\documentclass[11pt]{article}

\input epsf
\usepackage{latexsym}
\usepackage{amssymb}
\usepackage{amsmath}
\usepackage[dvips]{graphicx}

\begin{document}
\centerline{\Large \bf  Primordial magnetic seed fields from extra dimensions}

\vskip 2 cm

\centerline{Kerstin E. Kunze
\footnote{E-mail: kkunze@usal.es, Kerstin.Kunze@cern.ch} }

\vskip 0.3cm

\centerline{{\sl F\'\i sica Te\'orica, Universidad de Salamanca,}}
\centerline{{\sl Plaza de la Merced s/n, E-37008 Salamanca, Spain }}

\vskip 1.5cm

\centerline{\bf Abstract}
\vskip 0.5cm
\noindent
Dynamical extra dimensions break the conformal invariance 
of Maxwell's equations in four dimensions. 
A higher dimensional background with $n$ contracting 
extra dimensions and four expanding dimensions is 
matched to an effectively four dimensional standard radiation
dominated universe.
The resulting spectrum for the magnetic field is calculated
taking into account also the momenta along the extra dimensions.
Imposing constraints from observations an upper limit on the  
strength of magnetic seed fields is found. 
Depending on the number of extra dimensions, cosmologically 
interesting magnetic fields can be created.

\vskip 1cm

\section{Introduction}
Magnetic fields are ubiquitous in the universe.
Most galaxies, cluster of galaxies and even 
the Coma supercluster and radio galaxies at redshift
$z\simeq 2$ have been found to be endowed with
magnetic fields (for reviews see\cite{mag}-\cite{gr}). 

The average field strength of the 
interstellar magnetic field in our Galaxy has been 
observed to be $3-4 \mu$G. Spiral galaxies in general
seem to have magnetic fields with strength
of the order of $10 \mu$G. The structure of these magnetic fields
is determined by a large scale component with a coherence length
of the order of the size of the visible disk and a small-scale
component of tangled fields.
There are a few spiral galaxies with exceptionally strong magnetic 
fields of the order of $50 \mu$G, which also have a 
very high star formation rate \cite{klein,widr}.

Magnetic fields associated with elliptical galaxies 
have field  strengths comparable to those observed in 
spiral galaxies. However, their structure seems to be
quite distinct from that found for magnetic fields
in spiral galaxies. The coherence length is much smaller
than the corresponding galactic scales and the structure 
appears to be random.

In clusters of galaxies, magnetic fields of strength of the 
order of upto a few $\mu$G are found
in the intracluster medium. The cluster center regions 
indicate strong magnetic fields with typical 
field strengths of the order of $10-30 \mu$G and in exceptional
cases upto $70 \mu$G \cite{eil,gr}.
The coherence length of the magnetic fields is of the order
of the scale of the cluster galaxies.

There is also evidence for the existence of magnetic fields
in structures on supercluster scale.
The Coma-Abell 1367 supercluster is observed to have 
a magnetic field of strength $0.2-0.6 \mu$G \cite{kim,widr}.
Finally, observations indicate the existence of 
magnetic fields in redshift $z\simeq 2$ radio galaxies \cite{widr}.

There is no direct observational evidence of magnetic fields
that are not associated with any collapsing or virialized 
structure \cite{widr}. However, it is possible to put upper
bounds on the strength of such cosmological magnetic fields
from anisotropy measurements of the cosmic microwave background
\cite{cmb} 
and from the abundances of light elements predicted by
standard big bang nucleosynthesis \cite{bbn}.

To explain the widespread existence of 
large scale magnetic fields in
the universe it is commonly assumed that a tiny 
magnetic seed field at the epoch of galaxy formation is 
amplified by a dynamo mechanism to its present strength
of a few microgauss in our Galaxy \cite{mag}-\cite{gr}.
The dynamo amplifies an initial seed magnetic field
exponentially. The amplification factor depends on 
the growth rate for the dominant mode of the dynamo 
and the amount of time during which the dynamo 
operates. In a flat universe with no cosmological
constant the initial seed magnetic field needs to have at 
least a field strength of the order of $10^{-20}$ G
to explain the current $\mu$G galactic field 
today \cite{rees}, \cite{mag}-\cite{gr}.
However, as it was pointed out in \cite{dlt}, this bound 
depends on the cosmological model. In a flat universe
with non-vanishing cosmological constant the lower
limit on the required initial magnetic field strength
can be lowered significantly. For reasonable cosmological
parameter the required strength of the initial
seed magnetic field is of the order of $10^{-30}$G.

There are different proposals for the origin of the 
magnetic seed field \cite{mag}-\cite{gr}.
A class of proposed models involves the creation of 
magnetic seed fields during an inflationary stage of the
very early universe \cite{tw,infl}.
In order to produce a magnetic seed field of 
significant strength the conformal invariance of 
Maxwell's equations has to be broken, for example, by 
gravitational couplings of the photon \cite{tw}.

The conformal invariance of Maxwell's equations 
in four dimensions can also be broken if an embedding into
a higher dimensional space-time with time-varying extra 
spatial dimensions is considered.
In relation with the creation of seed magnetic fields this
was first investigated in \cite{giov}.
It is assumed that the $D$ dimensional space-time can be written
as a direct product of a three dimensional space and an 
$n$ dimensional space. Vacuum space-times of this type are 
provided by the Kasner solutions, which in general admit
two classes of solutions: either 
expanding three dimensions and collapsing extra dimensions or
vice versa.
The higher dimensional background with dynamical extra dimensions
is matched to a standard four dimensional radiation dominated universe
with static extra dimensions.
In \cite{giov} it was found that magnetic fields of cosmologically
interesting strength can be generated only in the case
of contracting three dimensions and 
growing extra dimensions. 
The novel feature of the model under consideration here 
is that momenta along the extra dimensions are 
also taken into account. The final spectrum is obtained by integrating
over these internal momenta \cite{int,ks}. 
This leads to the generation 
of magnetic seed fields of cosmologically interesting strength 
in the case of expanding three dimensions and contracting 
extra dimensions. Imposing bounds from observations 
an upper bound on the strength of the magnetic field can be found.

Models with extra dimensions arise naturally in string/ M-theory
which also led to the possibility of large extra dimensions \cite{hw}.
In higher dimensional gravity the four dimensional Planck scale $M_4$
is no longer fundamental, instead the higher dimensional
Planck scale $M_D$ becomes the fundamental scale. With the assumption that 
the $n$ extra dimensions have a characteristic size $R$, using
Gauss' law, the $D$-dimensional and the four-dimensional Planck masses
$M_4$ and $M_D$, respectively, are related by \cite{md}
\begin{eqnarray}
M_4^2=R^nM_D^{n+2}.
\end{eqnarray} 
Experiments show that Newtonian gravity is valid at least down to 
length scales of the order of 1 mm \cite{1mm}. This implies 
a lower bound on the ratio $M_D/M_4$.

\section{Magnetic fields from extra dimensions}

The background space-time is assumed to be homogeneous and
anisotropic with a line element,
\begin{eqnarray}
ds^2=a^2(\eta)\left[d\eta^2-\delta_{ij}dx^idx^j\right]
-b^2(\eta)\delta_{AB}dy^Ady^B,
\end{eqnarray}
where $i,j=1,..,3$ and $A,B=4,..,3+n$, $n\geq 1$.
$a(\eta)$ and $b(\eta)$ are the scale factor 
of the external, 3-dimensional space and the 
internal, $n$-dimensional space, respectively.

It is assumed that for $\eta<-\eta_1$ both scale-factors are functions 
of time. At $\eta=-\eta_1$ this is matched to a radiation dominated
four dimensional flat universe with static extra dimensions, $b(\eta)=const.$. 
The solutions are given by 
\begin{eqnarray}
a(\eta)&=&a_1\left(-\frac{\eta}{\eta_1}\right)^{\sigma},\hspace*{1.5cm}
b(\eta)=b_1\left(-\frac{\eta}{\eta_1}\right)^{\lambda},\hspace*{0.8cm}
{\rm for} \hspace*{0.5cm} \eta<-\eta_1
\label{b1}\\
a(\eta)&=&a_1\left(\frac{\eta+2\eta_1}{\eta_1}\right),\hspace*{1.05cm}
b(\eta)=b_1, \hspace*{2.4cm}
{\rm for} \hspace*{0.5cm} \eta\geq-\eta_1
\label{b2}
\end{eqnarray}
In the following we set $a_1=1=b_1$.

For $\eta<-\eta_1$ the solution is given by the vacuum Kasner metric,
which determines the exponents $\sigma$ and $\lambda$ as functions 
of the number of extra dimensions $n$.
These are related to 
the Kasner exponents $\alpha_E$ and $\alpha_I$, satisfying the 
Kasner conditions $3\alpha_E+n\alpha_I=1$ and $3\alpha_E^2+n\alpha_I^2=1$,
by
\begin{eqnarray}
\sigma=\frac{\alpha_E}{1-\alpha_E},\hspace{3cm}
\lambda=\frac{\alpha_I}{1-\alpha_E}.
\end{eqnarray}
In the case of an expanding, external space and a contracting, internal
space the exponents $\sigma$ and $\lambda$ are of the form \cite{giov},
\begin{eqnarray}
\sigma=-\frac{1}{2}\left(\sqrt{\frac{3n}{n+2}}-1\right),\hspace{2cm}
\lambda=\sqrt{\frac{3}{n(n+2)}}.
\end{eqnarray}

Maxwell's equations in $D$ dimensions are given by
$\nabla_{\tilde{A}}F^{\tilde{A}\tilde{B}}=0$
with $F_{\tilde{A}\tilde{B}}=\nabla_{[\tilde{A}}A_{\tilde{B}]}$, 
$\tilde{A}, \tilde{B}=0,..,n+3$.
Here the interest is the electromagnetic field in the 
(3+1)-dimensional space-time. Thus it is assumed that 
$A_i=A_i(x^i,y^B,\eta)$ and $A_B=0$.
Using the radiation gauge
$A_0=0$, $\nabla_iA^i=0$,  Maxwell's equations imply 
\begin{eqnarray}
-\frac{1}{b^n}\partial_0\left[b^n\partial_0 A_i\right]
+\sum_{j=1}^3\partial_j\partial_j A_i
+\left(\frac{a}{b}\right)^2\sum_{B=4}^{3+n}\partial_B
\partial_BA_i=0,
\end{eqnarray}
where $\partial_0\equiv\frac{\partial}{\partial\eta}$,
$\partial_i\equiv\frac{\partial}{\partial x^i}$ and
$\partial_B\equiv\frac{\partial}{\partial y^B}$.

Furthermore, the canonical field $\Psi_i=b^{\frac{n}{2}}A_i$ is introduced
and the following expansion is used \cite{giov}
\begin{eqnarray}
\Psi_i(\eta,x^i,y^A)=\int\frac{d^3 kd^nq}{(2\pi)^{\frac{3+n}{2}}}
\sum_{\alpha}e^{\alpha}_i({\bf l})\left[
a_{l,\alpha}\Psi_l(\eta)e^{i{\bf l}\cdot{\bf X}}
+a^{\dagger}_{-l,\alpha}\Psi^{*}_l(\eta)e^{-i{\bf l}\cdot{\bf X}}
\right],
\end{eqnarray}
where $l^{\mu}$ is a $(3+n)-$vector with components
$l^i\equiv k^i$, $l^A\equiv q^A$. Moreover, ${\bf l}\cdot{\bf X}=
{\bf k}\cdot{\bf x}+{\bf q}\cdot{\bf y}$.
$\alpha$ runs over the polarizations.
In the background (\ref{b1}), $\eta<-\eta_1$, this results in the mode equation
\begin{eqnarray}
\Psi_l''+\left[k^2+\left(-\frac{\eta}{\eta_1}\right)^{2\beta}
q^2-\frac{N}{\eta^2}\right]\Psi_l=0,
\label{psi}
\end{eqnarray}
where $'\equiv\frac{\partial}{\partial\eta}$ and
$N\equiv\frac{1}{4}\left(n\lambda-1\right)^2-\frac{1}{4}$.
Furthermore, $\beta\equiv\sigma-\lambda$.
$\beta<0$ since only solutions with contracting extra dimensions
will be discussed. $-1\leq\beta<-1/(1+\sqrt{3})$, where the 
lower boundary corresponds to $n=1$ and the upper bound gives the 
value for large $n$.

Equation (\ref{psi}) can be solved in a closed form for 
one extra dimension $n=1$. In general, for $n>1$, to our knowledge, 
apart from the case $n=6$, there are no solutions in closed form. 
However, it is possible to find approximate solutions.

\begin{itemize}
\item[-]
For $n=1$ and $\eta<-\eta_1$ 
the equation for $\Psi_l$ (cf. equation (\ref{psi})) reads
\begin{eqnarray}
\Psi_l''+\left[k^2+\left(-\frac{\eta}{\eta_1}\right)^{-2}q^2+
\frac{1}{4\eta^2}\right]\Psi_l=0,
\end{eqnarray}
which is solved by
\begin{eqnarray}
\Psi_l=\frac{\sqrt{\pi}}{2}e^{\frac{\pi}{2}q \eta_1}
\frac{\left(-k\eta\right)^{\frac{1}{2}}}{\sqrt{k}}
H_{i q\eta_1}^{(2)}(-k\eta),
\label{p-n1}
\end{eqnarray}
satisfying the Wronskian condition $\Psi_l^{* \prime}\Psi_l-\Psi_l'\Psi_l^{*}=i$
and $H_{\nu}^{(2)}(z)$ is the Hankel function of the second kind.

\item[-]
For $n>1$ and $\eta<-\eta_1$, in general approximate solutions can be found
to the mode equation (\ref{psi}).
In this case, there is a natural distinction into two
cases \cite{int,ks}.

\begin{enumerate}
\item
For $\left(-\frac{\eta}{\eta_1}\right)^{2\beta}q^2<k^2$,
or $\omega_q<\omega_k$
in terms of the physical frequencies $\omega_k=k/a(\eta)$
and $\omega_q=q/b(\eta)$, 
equation (\ref{psi}) becomes approximately,
\begin{eqnarray}
\Psi_l''+\left[k^2-\frac{N}{\eta^2}\right]\Psi_l=0,
\end{eqnarray}
which is solved by
\begin{eqnarray}
\Psi_l=\frac{\sqrt{\pi}}{2}\frac{\sqrt{-k\eta}}{\sqrt{k}}H_{\mu}^{(2)}(-k\eta),
\label{psi-mult1}
\end{eqnarray}
where $H_{\mu}^{(2)}$ is the Hankel function of the second kind and
$\mu^2\equiv\frac{1}{4}+N\Rightarrow \mu=\frac{1}{2}(n\lambda-1)$. 
The mode functions satisfy the Wronskian condition.

\item
For $\left(-\frac{\eta}{\eta_1}\right)^{2\beta}q^2>k^2$, or
$\omega_q>\omega_k$,
equation (\ref{psi}) can be approximated by,
\begin{eqnarray}
\Psi_l''+\left[\left(-\frac{\eta}{\eta_1}\right)^{2\beta}q^2-\frac{N}{\eta^2}
\right]\Psi_l=0,
\end{eqnarray}
which is solved by 
\begin{eqnarray}
\Psi_l=\frac{\sqrt{\pi}}{2}\left(-\kappa\eta\right)^{\frac{1}{2}}
H_{\mu\kappa}^{(2)}\left[\left(-q\eta\right)\kappa\left(-\frac{\eta}{\eta_1}
\right)^{\beta}\right],
\label{psi-mult2}
\end{eqnarray}
where $\kappa\equiv\frac{1}{\beta +1}$ and $\mu=\frac{1}{2}(n\lambda-1)$.

\end{enumerate}
The case $q=0$ is covered by the first case, 
$\left(-\frac{\eta}{\eta_1}\right)^{2\beta}q^2<k^2$, thus the solutions are not
written explicitly.

\end{itemize}

In the background (\ref{b2}), for $\eta\geq -\eta_1$, the mode equation is given by
\begin{eqnarray}
\Psi_l''+\left[k^2+\left(\frac{\eta+2\eta_1}{\eta_1}\right)^2q^2
\right]\Psi_l=0.
\end{eqnarray}
Introducing $z\equiv\left(\frac{2q}{\eta_1}\right)^{\frac{1}{2}}
\left(\eta+2\eta_1\right)$ and $\alpha\equiv -\frac{\eta_1 k^2}{2q}$
this can be transformed into the equation  for parabolic
cylinder functions \cite{giov2},
\begin{eqnarray}
\frac{d^2\Psi_l}{dz^2}+\left[\frac{z^2}{4}-\alpha\right]\Psi_l=0,
\end{eqnarray}
which is solved by
\begin{eqnarray}
\Psi_l=\frac{1}{\sqrt{2}}\left(\frac{\eta_1}{2q}\right)^{\frac{1}{4}}
\left[c_{-}E(\alpha, z)+c_{+}E^*(\alpha, z)\right],
\label{psi-rad}
\end{eqnarray}
where the Wronskian condition on the mode functions was applied and the 
normalization for the Bogoliubov coefficients $|c_+|^2-|c_{-}|^2=1$
was used. Using the approximations ((19.24) of \cite{AS}) 
gives expressions for $\Psi_l$ and $\Psi_l'$ at $\eta=-\eta_1$. 
\begin{enumerate}
\item
Namely,
for $\omega_q/\omega_k<1$, it is found 
that 
\begin{eqnarray}
\Psi_l(-\eta_1)&\sim&\frac{1}{\sqrt{2k}}
[c_{-}e^{ik\eta_1+i\frac{\pi}{4}}+c_{+}e^{-ik\eta_1-i\frac{\pi}{4}}]\nonumber\\
\Psi_l'(-\eta_1)&\sim&-\sqrt{\frac{k}{2}} 
[c_{-}e^{ik\eta_1-i\frac{\pi}{4}}+c_{+}e^{-ik\eta_1+i\frac{\pi}{4}}].
\end{eqnarray}

\item
For $\omega_q/\omega_k>1$ it follows that 
\begin{eqnarray}
\Psi_l(-\eta_1)&\sim&\frac{1}{\sqrt{2q}}
[c_{-}e^{i\frac{q\eta_1}{2}+i\frac{\pi}{4}}+
c_{+}e^{-i\frac{q\eta_1}{2}-i\frac{\pi}{4}}]
\nonumber\\
\Psi_l'(-\eta_1)&\sim&-\sqrt{\frac{q}{2}}
[c_{-}e^{i\frac{q\eta_1}{2}-i\frac{\pi}{4}}+
c_{+}e^{-i\frac{q\eta_1}{2}+i\frac{\pi}{4}}].
\end{eqnarray}
\end{enumerate}

The total magnetic energy density is given 
by \cite{ks}
\begin{eqnarray}
\rho=2\frac{R^n}{(2\pi)^{n+3}}\int \left[\left(\frac{k}{a}
\right)^2+\left(\frac{q}{b}\right)^2\right]^{\frac{1}{2}}
|c_{-}|^2 dV,
\end{eqnarray}
where, assuming that the volume consists of two spheres,
$dV=\frac{1}{a^3b^n}\frac{2\pi^{\frac{3}{2}}}{\Gamma(\frac{3}{2})}
k^2dk\wedge\frac{2\pi^{\frac{n}{2}}}{\Gamma(\frac{n}{2})}q^{n-1}dq$.
At $\eta=-\eta_1$ the comoving wavenumbers $k$ and $q$ are equal 
to the physical momenta, since $a_1=1=b_1$.
The spectral energy density $\rho(\omega_k)=d\rho/d {\rm log}\omega_k$
is then given by
\begin{eqnarray}
\rho(\omega_k)=16\frac{R^n}{(2\pi)^{n+3}}\frac{\pi^{1+\frac{n}{2}}}
{\Gamma(\frac{n}{2})}
\omega_k^{4+n}\int dY [1+Y^2]^{\frac{1}{2}}Y^{n-1}|c_{-}|^2,
\end{eqnarray}
where $Y\equiv\frac{\omega_q}{\omega_k}$, and $\omega_k=\frac{k}{a}$,
$\omega_q=\frac{q}{b}$.

During most of its history the universe had a very high conductivity,
implying that a primordial magnetic field evolves while its flux is conserved.
This makes the dimensionless ratio $r\equiv\rho_B/\rho_{\gamma}$ 
approximately constant \cite{tw}, where $\rho_B$ is the 
magnetic field energy density and $\rho_{\gamma}$ is the energy density
of the background radiation. Thus $r$ is a good measure of the strength
of a cosmological magnetic field. 
Furthermore, $r=\Omega_{em}/\Omega_{\gamma}$, where $\Omega=\rho/\rho_c$
with $\rho_c$ the critical energy density, and
$\Omega_{\gamma}=(H_1/H)^2(a_1/a)^4$.
Thus expressing the critical energy density in terms of the 
$D$-dimensional Planck mass $M_D$,
$\rho_c=\frac{3}{8\pi}R^nM_D^{n+2}H^2$, leads to
\begin{eqnarray}
r(\omega_k)=\frac{16}{3}\frac{8\pi}{(2\pi)^{n+3}}\frac{\pi^{1+\frac{n}{2}}}
{\Gamma(\frac{n}{2})}a^{-n}
\left(\frac{H_1}{M_D}\right)^{n+2}
\left(\frac{\omega_k}{\omega_1}\right)^{4+n}
\int_0^{Y_{max}} dY Y^{n-1}\left[1+Y^2\right]^{\frac{1}{2}}
|c_{-}|^2,
\label{rk}
\end{eqnarray}
where $\omega_1\equiv\frac{k_1}{a}$ and the maximal comoving wave number
$k_1\sim H_1$.
Furthermore, an upper cut-off $Y_{max}=\omega_{q_{max}}/\omega_k$ 
has been introduced.
This is justified by the sudden transition approximation, which 
is used here, since at the transition time, $\eta=-\eta_1$, the 
metric is continuous but not its first derivative. This means that 
for modes with periods much larger than the duration of the 
transition phase, the transition phase can be treated as
instantaneous. However, without an upper cut-off this 
type of approximation leads to an ultraviolet divergence \cite{div}.

For $q>0$ and $n=1$, that is one extra dimension,
continuously matching at $\eta=-\eta_1$ the solutions 
(\ref{p-n1}) and (\ref{psi-rad})
on superhorizon scales $k\eta_1\ll 1$, $q\eta_1\ll 1$
leads to the following 
Bogoliubov coefficients for $\omega_q/\omega_k<1$ and 
$\omega_q/\omega_k>1$,
\begin{eqnarray}
c_{-}e^{ik\eta_1}&\sim&
\frac{1}{\sqrt{2\pi}}\frac{1}{\sqrt{k\eta_1}}
\left[1+\frac{1}{2}\ln k\eta_1-ik\eta_1\ln k\eta_1\right]e^{-i\frac{\pi}{4}}
\hspace{2.2cm}{\rm for}\hspace{0.2cm}Y<1,\\
c_{-}e^{i\frac{q\eta_1}{2}}&\sim&
\frac{1}{\sqrt{2\pi}}\frac{1}{\sqrt{q\eta_1}}
\left[
1+\frac{1}{2}\ln k\eta_1-i q\eta_1\ln k\eta_1
\right]e^{-i\frac{\pi}{4}}
\hspace{2.2cm}{\rm for}\hspace{0.2cm}Y>1. 
\end{eqnarray}
Neglecting subleading terms, it follows that the 
ratio $r(\omega_k)$ is given by
\begin{eqnarray}
r(\omega_k)\sim
\frac{1}{3\pi^3}\left(\frac{H_1}{M_4}\right)^3
\left(\frac{M_5}{M_4}\right)^{-3}
\left(\frac{\omega_k}{\omega_1}\right)^3
\left(\ln \frac{\omega_k}{\omega_1}\right)^2
\frac{\omega_{q_{max}}}{\omega_1},
\label{rn1q}
\end{eqnarray}
where $\omega_{q_{max}}(\eta)=\frac{q_{max}}{b}$
and it was assumed that $\omega_{q_{max}}>\omega_k$.

The case $Y<1$ includes the limit $q=0$. Therefore together with 
$\rho_{em}(\omega_k)=2\frac{\omega_k^4}{\pi^2}|c_{-}(\omega_k)|^2$
the following expression for the ratio 
of magnetic to background radiation energy density
is obtained for $q=0$, $n=1$,
\begin{eqnarray}
r(\omega_k)\sim\frac{2}{3\pi^2}\left(\frac{H_1}{M_4}\right)^2
\left(\frac{\omega_k}{\omega_1}\right)^3
\left(\ln\frac{\omega_k}{\omega_1}\right)^2.
\label{rn1q0}
\end{eqnarray}

For more than one extra dimension, $n>1$,
the solutions for $\Psi_l$ and $\Psi_l'$ 
for $\eta<-\eta_1$ and $\eta>-\eta_1$
are matched  at $\eta=-\eta_1$ 
for $Y<1$ and  $Y>1$ for superhorizon modes, $k\eta_1\ll 1$,
$q\eta_1\ll 1$. 
This leads to the following expressions for the 
Bogoliubov coefficient $c_{-}$
\begin{eqnarray}
c_{-}e^{ik\eta_1}&\sim&\frac{2^{\mu-\frac{3}{2}}}{\sqrt{\pi}}\Gamma(\mu)
\left(k\eta_1\right)^{\frac{1}{2}-\mu}
\left[\left(\mu-\frac{1}{2}\right)\frac{1}{k\eta_1}+i\right]e^{-i\frac{\pi}{4}}
\hspace{2.2cm}
{\rm for} \hspace{0.5cm} Y<1,
\label{c1}
\\
c_{-}e^{i\frac{q\eta_1}{2}}&\sim&\frac{2^{\mu\kappa-\frac{3}{2}}}
{\sqrt{\pi}}\Gamma(\mu\kappa)\left(\kappa q\eta_1\right)^{\frac{1}{2}-\mu\kappa}
\left[\left(\mu-\frac{1}{2}\right)\frac{1}{q\eta_1}+i\right]e^{-i\frac{\pi}{4}}
\hspace{1.5cm}
{\rm for} \hspace{0.5cm} Y>1.
\label{c2}
\end{eqnarray}

Using the expressions for $|c_{-}|$ for $Y<1$ and $Y>1$, as provided by 
equations (\ref{c1}) and (\ref{c2}) for more than one extra dimension, $n>1$,
leads to the ratio of magnetic spectral energy density to background 
radiation density,
\begin{eqnarray}
r(\omega_k)&\sim& {\cal N}
a^{1+2\mu\kappa-n}
\left(\frac{H_1}{M_D}\right)^{n+2}
\left(\frac{\omega_{q_{max}}}{\omega_1}\right)^{n-2\mu\kappa}
\left(\frac{\omega_k}{\omega_1}\right)^3
\label{ro}
\end{eqnarray}
where 
\begin{eqnarray}
{\cal N}&\equiv& \frac{16}{3}\frac{8\pi}{(2\pi)^{n+3}}
\frac{\pi^{1+\frac{n}{2}}}{\Gamma(\frac{n}{2})}
\frac{2^{2\mu\kappa-3}}{\pi(n-2\mu\kappa)}\Gamma^2(\mu\kappa)
\kappa^{1-2\mu\kappa}
\left(\mu-\frac{1}{2}\right)^2
\nonumber
\end{eqnarray} 
where subleading terms have been omitted and 
$\omega_{q_{max}}>\omega_k$ was assumed.
The resulting spectrum is growing in frequency.

The expression for $q=0$ can be derived using the expression for
$c_{-}$  for $Y<1$
(cf. equation (\ref{c1})). 
Together with $\rho_{em}=2\frac{\omega^4}{\pi^2}|c_{-}|^2$
this implies for $q=0$, $n>1$
\begin{eqnarray}
r(\omega_k)\sim \frac{2^{n\lambda-2}}{3\pi^2}
\Gamma^2\left(\frac{n\lambda-1}{2}\right)
\left(2-n\lambda\right)^2\left(\frac{H_1}{M_4}\right)^2
\left(\frac{\omega_k}{\omega_1}\right)^{4-n\lambda}.
\label{roq0}
\end{eqnarray}
Furthermore, $n\lambda=\sqrt{\frac{3n}{n+2}}$.
Since $n\lambda<4$ the resulting spectrum
for $r(\omega_k)$ is increasing in frequency.

\section{Constraining the model}

The expressions for the ratio $r(\omega_k)$  
determining the ratio of the energy density of the magnetic field 
in comparison with the energy density of the background radiation 
contain several parameters apart from the physical frequencies 
$\omega_k$ and $\omega_{q_{max}}$.
The free parameters are the Hubble parameter at the time of transition 
$H_1$, the $D$-dimensional Planck mass $M_D$ and the number of extra 
dimensions $n$.  

There are several constraints from observations. 
$r(\omega_k)$ has to be less than one for all 
frequencies in order not to overclose the 
universe. For $r(\omega_k)$ increasing with 
frequency this implies $r(\omega_1)<1$.
This is the case for the spectra given by 
equations (\ref{ro}) and (\ref{roq0}) applicable for 
backgrounds with more than one extra dimension.
In the case of one extra dimension  
the expressions for $r(\omega_k)$ (cf.
equations (\ref{rn1q}) and (\ref{rn1q0}))
have a maximum at some frequency $\omega_2$.
Thus the constraint $r(\omega_2)<1$ is imposed.

Newtonian gravity has been tested down to
length scales of the order of 1 mm
\cite{1mm}. This implies the constraint
$\frac{M_D}{M_4}\geq (1.616\times 10^{-32})^{\frac{n}{n+2}}$.
Furthermore, with $T_1$ the temperature at the 
beginning of the radiation epoch, 
big bang nucleosynthesis requires that $T_1> 10$ MeV.
This imposes a bound on $H_1$ by using 
$\frac{H_1}{M_4}=1.66 g_{*}^{\frac{1}{2}}(T_1)\left(
\frac{T_1}{M_4}\right)^2$, where for $T_1>300$ GeV the number of effective
degrees of freedom is given by $g_{*}(T_1)=106.75$ \cite{kt},
namely, $\log\frac{H_1}{M_4}>-40.94$.

The ratio $r$ calculated at the galactic scale $\omega_G^{-1}$ of 
order of 1 Mpc
determines the strength of the primordial seed magnetic
field at the time of galaxy formation. 
In the standard picture of a galactic magnetic dynamo 
operating since the time of galaxy formation,
a seed magnetic field of at least
$B_s\sim 10^{-20}$G \cite{rees}, corresponding to 
$r(\omega_G)>10^{-37}$,
is needed to explain the currently
observed galactic magnetic field of a few $\mu$G \cite{rees}.
However, taking into account a non-vanishing cosmological 
constant, it was shown in \cite{dlt} that initial magnetic seed 
field strengths can be much below $10^{-20}$G. Thus $r(\omega_G)$
can be as low as $10^{-57}$ and correspondingly the magnetic seed
field $B_s\sim 10^{-30}$ G. 

In the following, using the constraint $r(\omega_1)<1$ or
$r(\omega_2)<1$, respectively, the constraint from
the size of the extra dimension and from big bang nucleosynthesis
an upper limit on the ratio $r(\omega_G)$ and thus the strength 
of the magnetic seed field strength at the time of galaxy formation
is derived. The strength of the seed field in terms of $r$ is given
by $B_s\sim 3 r^{\frac{1}{2}}\times 10^{-2}$ G \cite{tw}.

In addition, the maximally 
amplified frequency calculated with respect to present day 
$\omega_1(\eta_0)$ is given by 
$\omega_1\sim 6\times 10^{11}{\rm Hz}
\left(\frac{H_1}{M_4}\right)^{\frac{1}{2}}$ and 
the frequency corresponding to galactic scale,
$\omega_G\sim 10^{-14}$Hz \cite{giov}.
Furthermore, $r(\omega_G)$ is assumed to be 
of the form $r(\omega_G)=10^{-m}$ where the exponent $m$ will be 
constrained by  observational bounds. In the standard picture of the 
galactic dynamo, $m\leq 37$. 
In the following an upper bound on $-m$ will be found. 

For one extra dimension, $n=1$,
the spectra (\ref{rn1q}) and (\ref{rn1q0}) have a maximum at
$\frac{\omega_2}{\omega_1}=e^{-\frac{2}{3}}$. Thus the constraint
of the critical density is imposed by requiring $r(\omega_2)<1$.

In the case where the momenta lying in the extra dimension
are not taken into account, that is $q=0$,
$r(\omega_G)=10^{-m}$ where $\omega_G=10^{-14}$Hz
implies,
\begin{eqnarray}
-m=\log\frac{2}{3\pi^2}+\frac{1}{2}\log\frac{H_1}{M_4}
+3\log\frac{10^{-14}}{6\times 10^{11}}
+\log\left[\ln\frac{10^{-14}}{6\times 10^{11}}
-1.1513\log\frac{H_1}{M_4}\right]^2.
\end{eqnarray}
Big bang nucleosynthesis requires $\log\frac{H_1}{M_4}>-40.94$
and the constraint $r(\omega_2)<1$ implies $\log\frac{H_1}{M_4}<1.2$.
Evaluating $m$ at the upper limit   $\log\frac{H_1}{M_4}=1.2$
gives $r(\omega_G)<10^{-74}$
corresponding to a magnetic seed field
strength of $B_s<10^{-39}$ G. Thus magnetic fields
created in this setting are too weak in order to seed the 
galactic magnetic dynamo.

For $q>0$ and $n=1$ the various constraints mentioned above
applied to the expression for $r(\omega_k)$ (cf. equation (\ref{rn1q}))
lead to the constraint on $m$
\begin{eqnarray}
-m<\log\frac{9e^2}{4}+3\log\frac{10^{-14}}{6\times 10^{11}}-\frac{3}{2}
\log\frac{H_1}{M_4}
+\log\left[
\ln\frac{10^{-14}}{6\times 10^{11}}
-1.1513\log\frac{H_1}{M_4}
\right]^2.
\end{eqnarray}
Evaluating $m$ at the lower bound $\log\frac{H_1}{M_4}=-40.94$
results in the bound $r(\omega_G)<10^{-13}$ corresponding 
to a magnetic seed field strength of $B_s<10^{-8}$ G.
Thus in this case the lower bound on the magnetic seed field 
imposed by the galactic dynamo can be satisfied easily.

Assuming that $T_1\sim M_D$ results in an additional 
constraint on $\log\frac{H_1}{M_4}$ by using the 
bound on the size of the extra dimensions.
Namely, for any $n$,
\begin{eqnarray}
\log\frac{H_1}{M_4}>\log 17.15+\frac{2n}{n+2}\log(1.616\times 10^{-32}).
\label{h1}
\end{eqnarray}

This gives a bound on $\log\frac{H_1}{M_4}$ stronger than the 
one from big bang nucleosynthesis only upto three extra dimensions
$n\leq 3$.
In particular in the case at hand, for $n=1$, 
it implies $\log\frac{H_1}{M_4}>-19.96$. Evaluating 
$m$ at this value of $\log\frac{H_1}{M_4}$ leads to
$r(\omega_G)<10^{-43}$ and correspondingly
the magnetic seed field strength
$B_s<10^{-23}$ G. Thus in the case where $T_1\sim M_5$ 
the created magnetic seed field satisfies the weaker 
bound of $B_s>10^{-30}$ G.

For more than one extra dimension $n>1$ and $q>0$
the constraint on $r(\omega_k)$ (cf. equation (\ref{ro}))
at $\omega_1$ together with  $r(\omega_G)=10^{-m}$
leads to
\begin{eqnarray}
-m<-\frac{3}{2}\log\frac{H_1}{M_4}+3\log\frac{10^{-14}}{6\times 10^{11}}.
\end{eqnarray}
Using the constraint from big bang nucleosynthesis $\log\frac{H_1}{M_4}>
-40.94$ results in
$-m<-15.9$ and thus $r(\omega_G)<10^{-16}$ and hence
seed magnetic fields with strengths upto $B_s<10^{-10}$ G
can be created.
Assuming that the temperature at the beginning of the 
radiation epoch, $T_1$, is given by $M_D$, that is 
$T_1\sim M_D$, changes the bound on $m$ for two and 
three extra dimensions (cf. equation (\ref{h1})). 
In this case, for $n=2$ extra dimensions, $-m<-31.5$
implying $r(\omega_G)<10^{-32}$ and the magnetic field
strength $B_s<10^{-18}$ G.
For $n=3$ extra dimensions,  $-m<-21.95$ and hence
$r(\omega_G)<10^{-22}$ and the magnetic field strength
$B_s<10^{-13}$G.

\begin{figure}[ht]
\centerline{\epsfxsize=2.5in\epsfbox{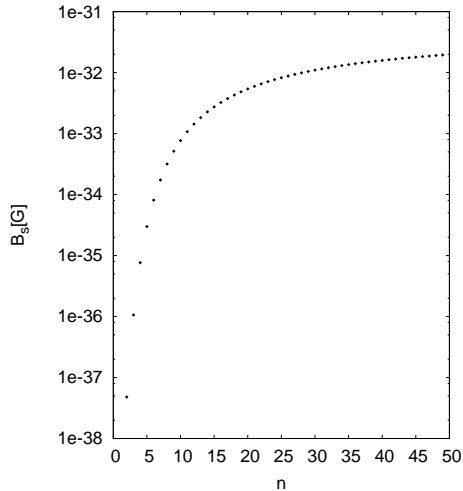}}
\caption{The graph shows the maximal
field strength for the 
seed magnetic field $B_s$ as a function of extra dimensions $n$, $n>1$,
for $q=0$, that is the internal momenta are not taken into account.
}
\label{fig}
\end{figure}

This is to be compared with the case where the internal
momenta are not taken into account \cite{giov}.
Applying the constraints to equation (\ref{roq0})
implies 
\begin{eqnarray}
-m<\left(1-\frac{n\lambda}{4}\right)
\log\left[\frac{2^{n\lambda-2}}{3\pi^2}
\Gamma^2\left(\frac{n\lambda-1}{2}\right)\left(2-n\lambda\right)^2
\right]+\left(4-n\lambda\right)\log\frac{10^{-14}}{6\times 10^{11}}.
\end{eqnarray}
In this case the bound on $-m$ depends on the number of extra dimensions
$n$. This is related to the fact that the spectral index in the 
expression for $r(\omega_k)$ (cf. equation (\ref{roq0}))
is given by $4-n\lambda$ and thus
depends explicitly on the number of dimensions. 
In the case, where $n>1$ and $q>0$, the spectral 
index is 3, independent of the number of extra dimensions.
In figure \ref{fig} the magnetic seed field 
strength $B_s$ is plotted as a function of the number of
extra dimensions $n$ in the case $n>1$, $q=0$.
As can be seen the resulting values for 
$B_s$ are very small, much below even the 
weaker constraint, $B_s>10^{-30}$G \cite{dlt}.

In the cases $n=1$ and $n>1$ for $q>0$,  $\omega_{q_{max}}=q_{max}/b$
appears as a parameter in the expressions for 
$r(\omega_k)$ (cf. equations (\ref{rn1q})
and (\ref{ro})). Assuming that $q_{max}\sim k_1$
leads to $\omega_{q_{max}}/\omega_1\sim a$. Using this 
in $r(\omega_2)<1$, for $n=1$, and in $r(\omega_1)<1$,
for $n>1$, leads in both cases to a constraint of the form
\begin{eqnarray}
{\cal N}_{*}a_0\left(\frac{H_1}{M_4}\right)^{n+2}\left(
\frac{M_D}{M_4}\right)^{-(n+2)}<1,
\label{md}
\end{eqnarray}
where ${\cal N}_{*}=\frac{4}{27e^2\pi^3}$ for $n=1$
and ${\cal N}_{*}={\cal N}$ for $n>1$.
If there are no additional constraints then 
equation (\ref{md}) implies a lower bound on $M_D/M_4$, which has to
be compared with the lower bound provided by the size of the
extra dimensions.
However, if in addition $T_1\sim M_D$ is imposed, then
equation (\ref{md}) together with $\frac{H_1}{M_4}\sim 1.66 
g_{*}^{\frac{1}{2}}\left(\frac{M_D}{M_4}\right)^2$ implies an 
upper bound on $M_D/M_4$, namely,
\begin{eqnarray}
\log\frac{M_D}{M_4}<-\frac{\log\left(3\times 10^{31}{\cal N}_{*}\right)}
{n+3}-\frac{n+\frac{5}{2}}{n+3}\log 1.66g_{*}^{\frac{1}{2}},
\end{eqnarray}
where $a_0\sim 3\times 10^{31}\left(\frac{H_1}{M_4}\right)^{\frac{1}{2}}$
was used. This bound is always larger than the lower bound on
$\log\frac{M_D}{M_4}$ provided by the size of the extra
dimensions. Thus, the assumption $T_1\sim M_D$ is consistent
with the various constraints.
Moreover, although this upper bound on $M_D/M_4$ leads to 
an upper bound on $H_1/M_4$, the maximal strength of 
the magnetic seed field is not changed, since 
for $n\geq 1, q>0$ this 
was evaluated at the lower boundary of $\log \frac{H_1}{M_4}$.

\section{Conclusions}

The origin of magnetic fields on galactic and extragalactic scales
is still an open problem. Different types of mechanisms have been
proposed. In particular, in \cite{giov} the creation of magnetic fields
due to dynamical extra dimensions was proposed.  Along these lines,
here, a model consisting of two phases has been investigated.
A higher dimensional epoch with three expanding, external (spatial)
dimensions and $n$ contracting, internal dimensions is 
matched to a standard radiation dominated phase with static
extra dimensions. Taking the internal momenta into account
the final expression for the ratio $r$ of magnetic field energy density to
background radiation energy density is obtained by integrating over the 
internal modes. In doing so the sudden approximation requires the 
introduction of a maximal frequency in the internal momentum space.
The resulting spectrum is constrained by imposing bounds
from observations, such as, the constraint from 
critical energy density, the size of the extra dimensions
and big bang nucleosynthesis. 

For one extra dimension, $n=1$, it was found that 
in the case where the momenta along the 
extra dimension are not taken into account, $q=0$,
only very weak magnetic seed fields are created, $B_s<10^{-39}$ G.
However, in the case $q>0$ magnetic seed fields 
as strong as $10^{-8}$ G can be obtained in general.
Imposing the additional constraint $T_1\sim M_5$ leads to
magnetic seed fields $B_s<10^{-23}$ G which satisfy
the lower bound in a $\Lambda$ universe \cite{dlt}.

In models with more than one extra dimension, $n>1$,
strong magnetic seed fields can be created if the 
internal momenta are taken into account. In particular,
not assuming that the temperature at the beginning of the 
radiation epoch is of the order of the $D$-dimensional
Planck scale allows for the creation of seed magnetic fields
with strengths of upto $10^{-10}$ G. For more than three
extra dimensions, this also holds if $T_1\sim M_D$
is assumed. With this assumption for two and three 
extra dimensions results in weaker magnetic seed fields,
with maximal field strengths, $B_s<10^{-18}$ G for two
extra dimensions and $B_s<10^{-13}$G for three extra
dimensions. 

Therefore, in this particular model with extra dimensions, taking 
into account the momenta along the extra dimensions  
allows for the creation of strong magnetic fields.

\section{Acknowledgements}

I would like to thank M. A. V\'azquez-Mozo for 
useful discussions.
This work has been supported by the programme
``Ram\'on y Cajal'' of the M.E.C. (Spain).
Partial support by Spanish Science Ministry
Grants FPA 2002-02037 
and BFM 2003-02121 is acknowledged.


\begin{thebibliography}{99}
\bibitem{mag}
P.P. Kronberg, Rep. Prog. Phys. {\bf 57} (1994) 325;
E. G. Zweibel and C. Heiles, Nature {\bf 385} (1997) 131;
M.~Giovannini,
Int.\ J.\ Mod.\ Phys.\ D {\bf 13} (2004) 391.

\bibitem{widr}
L. M. Widrow, Rev. Mod. Phys. {\bf 74} (2002) 775.

\bibitem{gr}
D. Grasso and H.R. Rubinstein, Phys. Rep. {\bf 348} (2001) 163.

\bibitem{klein}
U. Klein, R. Wielebinski and H. W. Morsi, Astron. Astrophys.
{\bf 190} (1988) 41.

\bibitem{eil}
J.~A.~Eilek,
in ``Diffuse Thermal and Relativistic Plasma in Galaxy Clusters''
Ed. by Hans Bohringer, Luigina Feretti, Peter Schuecker. Garching, Germany (1999).
(arXiv:astro-ph/9906485)


\bibitem{kim}
K.-T. Kim, P. P. Kronberg, G. Giovannini 
and T. Venturi,
Nature {\bf 341} (1989) 720.


\bibitem{cmb}
J.~D.~Barrow, P.~G.~Ferreira and J.~Silk,
Phys.\ Rev.\ Lett.\  {\bf 78} (1997) 3610;
R.~Durrer, T.~Kahniashvili and A.~Yates,
Phys.\ Rev.\ D {\bf 58} (1998) 123004;
K.~Subramanian and J.~D.~Barrow,
Phys.\ Rev.\ Lett.\  {\bf 81} (1998) 3575.


\bibitem{bbn}
B.~l.~Cheng, D.~N.~Schramm and J.~W.~Truran,
Phys.\ Rev.\ D {\bf 49} (1994) 5006;
B.~l.~Cheng, A.~V.~Olinto, D.~N.~Schramm and J.~W.~Truran,
Phys.\ Rev.\ D {\bf 54} (1996) 4714;
D.~Grasso and H.~R.~Rubinstein,
Phys.\ Lett.\ B {\bf 379} (1996) 73;
P.~J.~Kernan, G.~D.~Starkman and T.~Vachaspati,
Phys.\ Rev.\ D {\bf 54} (1996) 7207.


\bibitem{rees}
M. J. Rees, Q. J. R. Astron. Soc. {\bf 28} (1987) 197. 

\bibitem{dlt}
A.~C.~Davis, M.~Lilley and O.~Tornkvist,
Phys.\ Rev.\ D {\bf 60} (1999) 021301.



\bibitem{tw}
M.~S.~Turner and L.~M.~Widrow,
Phys.\ Rev.\ D {\bf 37} (1988) 2743;

\bibitem{infl}
B.~Ratra,
Astrophys.\ J.\  {\bf 391} (1992) L1;
D.~Lemoine and M.~Lemoine,
Phys.\ Rev.\ D {\bf 52} (1995) 1955;
M.~Gasperini, M.~Giovannini and G.~Veneziano,
Phys.\ Rev.\ Lett.\  {\bf 75} (1995) 3796.


\bibitem{giov}
M.~Giovannini,
Phys.\ Rev.\ D {\bf 62} (2000) 123505.


\bibitem{int}
R.~Durrer and M.~Sakellariadou,
Phys.\ Rev.\ D {\bf 62} (2000) 123504;
R.~Durrer, K.~E.~Kunze and M.~Sakellariadou,
New Astron.\ Rev.\  {\bf 46} (2002) 659.

\bibitem{ks}
K.~E.~Kunze and M.~Sakellariadou,
Phys.\ Rev.\ D {\bf 66} (2002) 104005.


\bibitem{hw}
P.~Ho\u{r}ava and E.~Witten,
Nucl.\ Phys.\ B {\bf 460} (1996) 506

\bibitem{md}
N.~Arkani-Hamed, S.~Dimopoulos and G.~R.~Dvali,
Phys.\ Lett.\ B {\bf 429} (1998) 263;
I.~Antoniadis, N.~Arkani-Hamed, S.~Dimopoulos and G.~R.~Dvali,
Phys.\ Lett.\ B {\bf 436} (1998) 257;
V.~A.~Rubakov,
Phys.\ Usp.\  {\bf 44} (2001) 871
[Usp.\ Fiz.\ Nauk {\bf 171} (2001) 913].

\bibitem{1mm}
C.~D.~Hoyle, U.~Schmidt, B.~R.~Heckel, E.~G.~Adelberger, J.~H.~Gundlach, 
D.~J.~Kapner and H.~E.~Swanson,
Phys.\ Rev.\ Lett.\  {\bf 86} (2001) 1418.


\bibitem{giov2}
M.~Giovannini,
Phys.\ Rev.\ D {\bf 55} (1997) 595.


\bibitem{AS}
{\it Handbook of Mathematical Functions}, ed. by 
M. Abramowitz and I. A. Stegun (Dover Publications, New York, 1965).


\bibitem{div} 
B.~L.~Hu,
Phys.\ Rev.\ D {\bf 9} (1974) 3263;
J.~Garriga and E.~Verdaguer,
Phys.\ Rev.\ D {\bf 39} (1989) 1072;
M.~R.~de Garcia Maia and J.~D.~Barrow,
Phys.\ Rev.\ D {\bf 50} (1994) 6262.


\bibitem{kt}
E.~W.~Kolb and M.~S.~Turner,
{\it The Early Universe},
 Redwood City, USA: Addison-Wesley (1990).

\end{thebibliography}
\end{document}